# Magnetic twisting in an artificial ferrimagnet: Anisotropic magnetoresistance on Py/Gd/Py/Gd/Py/SiN$_x$ multilayers

Kai Zhang[1], Y. X. Niu[1], Yang Meng[2,3*], Hong-Wu Zhao[2,3,4] & J. Li[1*]

[1]*International Center for Quantum Materials, School of Physics, Peking University, Beijing 100871, China*

[2]*Beijing National Laboratory for Condensed Matter Physics, Institute of Physics, Chinese Academy of Sciences, Beijing 100190, China*

[3]*School of Physical Sciences, University of Chinese Academy of Sciences, Beijing 100049, China*

[4]*Songshan Lake Materials Laboratory, Dongguan 523808, China*

*Correspondence to: ymeng@iphy.ac.cn, jiali83@pku.edu.cn

The intensive study of non-collinear magnets promotes an urgent demand for the quantitative characterization of the non-collinear magnetic structures, which host numerous exotic phenomena. Here we systematically study the non-collinear magnetic structure of an artificial ferrimagnetic multilayer. The AMR measurements reveal two distinct twisted states whose magnetic structures can be quantitatively characterized with the assistance of micromagnetic simulations. Our results manifest AMR as an ideal probe of the non-collinear magnetic structure in artificial ferrimagnets.

## I. INTRODUCTION

Spin-dependent phenomena hosted by collinear magnets (e.g. ferromagnets or antiferromagnets) have been successfully harnessed in diverse magnetic logic and data storage devices in the past years [1]. Recently, a wealth of non-collinear magnets that could be stabilized by Dzyaloshinskii-Moriya interaction [2], dipolar interaction [3,4], or higher-order exchange interactions [5] were discovered. A wide range of exotic phenomena has emerged in numerous non-collinear magnets, such as large anomalous Hall effect [6,7], vector spin Seebeck effect [8], ultrafast optical parametric pumping [9], chiral domain walls' damping [10], etc. To detect non-collinear magnetization in these systems, sophisticated experimental techniques such as Lorentz transmission electron microscopy [11] or x-ray magnetic linear dichroism [12], are utilized. In particular, our recent work demonstrates that the switching of magnon chirality becomes feasible hinging on the emergence of the non-collinear magnetic phase (i.e. twisted state) in an artificial ferrimagnet [13]. New functionalities and computing architectures based on this artificial ferrimagnet may open fascinating perspectives for novel spintronics applications [14]. Hitherto, the characterization of the non-collinear magnetic structure in such systems has consistently depended on the detection of the nonlinearity in magnetization curves [15], which is a non-electric (off-chip) detection. Although the non-collinear magnetic phase of such artificial ferrimagnets (hereafter referred to as the twisted state) has been well studied by conventional magnetization-curve measurement, electrical (on-chip) detection of the twisted state is particularly desirable for quantitative purpose and potential applications.

In this work, we systematically study the anisotropic magnetoresistance (AMR) in an artificial ferrimagnetic Py/Gd/Py/Gd/Py/SiN$_x$ multilayer. The transformation from the collinear state to the twisted state can be explicitly identified by AMR measurements. The AMR results at low temperature resolve two distinct magnetic windings in the twisted states (occurring at different threshold fields), which can be assigned to surface twisting and bulk twisting in the multilayer. We also note a good agreement between the AMR results and micromagnetic simulations. Our results manifest AMR as a sensitive probe for the non-collinear magnetic structure of artificial ferrimagnets. Utilizing AMR measurements, the complex non-collinear magnetic profile could be investigated quantitatively in an electrical (on-chip) manner.

## II. EXPERIMENT AND SIMULATION

The Py(2.5)/Gd(3)/Py(2.5)/Gd(3)/Py(2.5)/SiN$_x$ multilayer sample was deposited on single crystalline Al$_2$O$_3$(0001) substrates by DC magnetron sputtering under an Ar pressure of 3.5 mTorr at room temperature, the numbers in parenthesis are thicknesses in the unit of nanometers. To obtain alternate layers of Py and Gd with different thicknesses, high purity Py (99.95%) and Gd (99.9%) targets were sputtered for different durations in sequence. The deposition rates were 2.4 nm/min and 1.2 nm/min for Gd and Py, respectively. The SiN$_x$ capping layer (12 nm) was deposited on top of the sample to protect it from oxidation.

The quality of the multilayer sample is characterized by energy dispersive X-ray spectroscopy (EDS). Figure 1 (a) to (d) show the EDS maps of elements Gd, Ni, Fe and Al. The EDS maps of Ni and Gd illustrate the continuous Py and Gd layers with uniform elemental distributions and clear boundaries between Py and Gd. The quality of the sample is also characterized by the low-angle x-ray reflectivity (XRR). As shown in Fig. 1 (e), the periodical oscillations observed in the XRR scans confirm the well-defined interfaces of the multilayer sample, in consistent with the result of EDS mapping. The fitting of XRR reveals a smooth Py/Gd interface with an RMS roughness of 0.5 nm, which promotes a strong antiferromagnetic interfacial coupling at the Py/Gd and Gd/Py interfaces [16].

Static magnetization of the sample was investigated in the temperature range of 10–300 K in magnetic fields up to 70 kOe using a conventional Quantum Design Magnetic Property Measurement System (MPMS) SQUID magnetometer. The magnetic properties of the substrate were measured separately, and its contribution was subtracted from the total magnetic moment of the sample.

The samples were patterned into standard Hall bar with a length of L = 4 mm and a width of w = 100 μm, by optical lithography and ion beam etching. Transport measurements were conducted using a conventional four-probe technique with an electrical current $I$ (0.1 mA) flowing in the film plane. The magnetoresistance (MR) was recorded with a sweeping magnetic field parallel to or perpendicular to the Hall bar in a physical property measurement system (Quantum Design PPMS-9T system). The anisotropic magnetoresistance (AMR) was measured with a rotatable sample stage. A static magnetic field was kept in the film plane during AMR measurements.

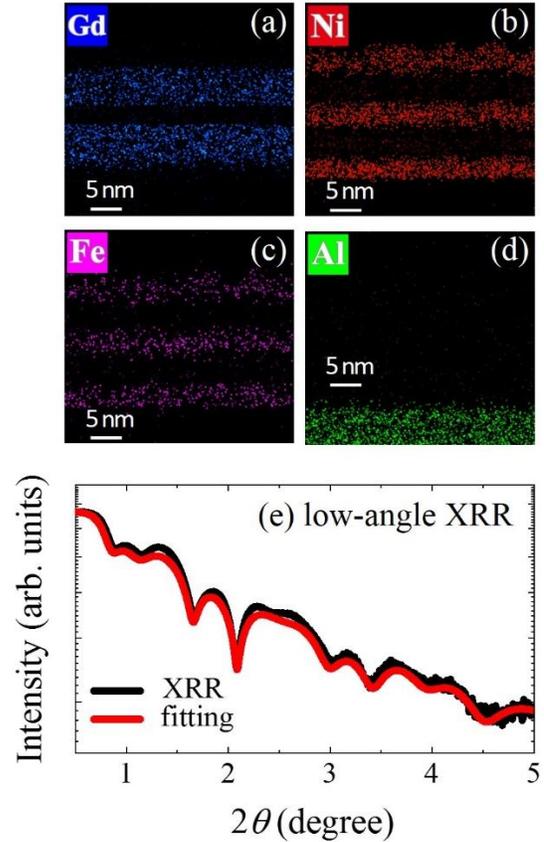

Fig. 1. Elemental distributions in Py/Gd/Py/Gd/Py/SiN$_x$ multilayer sample measured by energy dispersive X-ray spectroscopy (EDS). EDS maps of (a) element Gd, (b) element Ni, (c) element Fe and (d) element Al. (e) Low-angle x-ray reflectivity (XRR) scan of the multilayer sample.

In order to examine the magnetization reversal mechanism of the multilayer sample, we reproduced the hysteresis loops at different temperatures by the micromagnetic simulation using the Object Oriented MicroMagnetic Framework (OOMMF) code based on the Landau-Lifshitz-Gilbert equation [17]. The magnetic structure of Py/Gd multilayer in the twisted state is a one-dimensional spin chain winding along the film normal direction. To imitate such one-dimensional spin chain, the cell size was chosen to be 100 μm×100 μm×0.25 nm. Within each 100 μm×100 μm plane, only one spin was taken into consideration. Thus the whole system behaves like a one-dimensional spin chain.

## III. RESULTS

The Py/Gd/Py/Gd/Py multilayer (hereafter simplified as the Py/Gd multilayer) is an artificial ferrimagnet with the strong interfacial antiferromagnetic coupling between Py moment $M_{Py}$ and Gd moment $M_{Gd}$ [13]. The distinct temperature dependences of $M_{Py}$ and $M_{Gd}$ result in a compensation temperature $T_M$ where the staggered $M_{Py}$ and $M_{Gd}$ are fully compensated, leading to a zero macroscopic magnetization [18]. As described in the literatures [13,19], $M_{Py}$ governs the macroscopic magnetization of the Py/Gd multilayer for $T > T_M$, referred to as Py-aligned phase. And $M_{Gd}$ dominates the macroscopic magnetization for $T < T_M$ (i.e. Gd-aligned phase). For both scenarios, the subordinate moment (slave) aligns antiparallel to the magnetic field $H$ when the dominant moment (master) is aligned with $H$, as a result of the strong interfacial antiferromagnetic coupling. In addition, a magnetic twisted state could be introduced in a sufficiently strong magnetic field [20].

We first characterize the Py/Gd multilayer sample by the conventional measurements of magnetization curves in an in-plane sweeping $H$. Figure 2 (a) and (b) present the positive half branches of the magnetization curves at $T = 90$ K and $T = 10$ K, respectively. The different remanences extracted in Fig. 2 (a) and (b) suggest a temperature-sensitive macroscopic magnetization of the Py/Gd multilayer. The temperature-dependent in-plane magnetization at $H = 50$ Oe exhibits a local minimum at $T = 67$ K [Fig. 2 (e)], revealing that the compensation temperature $T_M$ is 67 K for the Py/Gd multilayer sample. Note that the magnetization curves rise nonlinearly with $H$ in Fig. 2 (a) and (b). The bending of the magnetization curves is observed at a threshold field [grey dotted lines in Fig. 2 (a) and (b)], indicating the formation of the non-collinear magnetic structure in the Py/Gd multilayer. Below the threshold field, the Py/Gd multilayer behaves as a ferromagnet. Slave moment is antiferromagnetically coupled to master moment and always antiparallel to $H$. The Zeeman energy of slave moment due to $H$ is negligible in this scenario. On the contrary, above the threshold field, the Zeeman energy could become non-negligible and introduce a depth-dependent non-collinear magnetic profile in the Py/Gd multilayer, which is usually referred to as the twisted state [20]. The threshold field of the twisted state is called the twisting field $H_{twist}$.

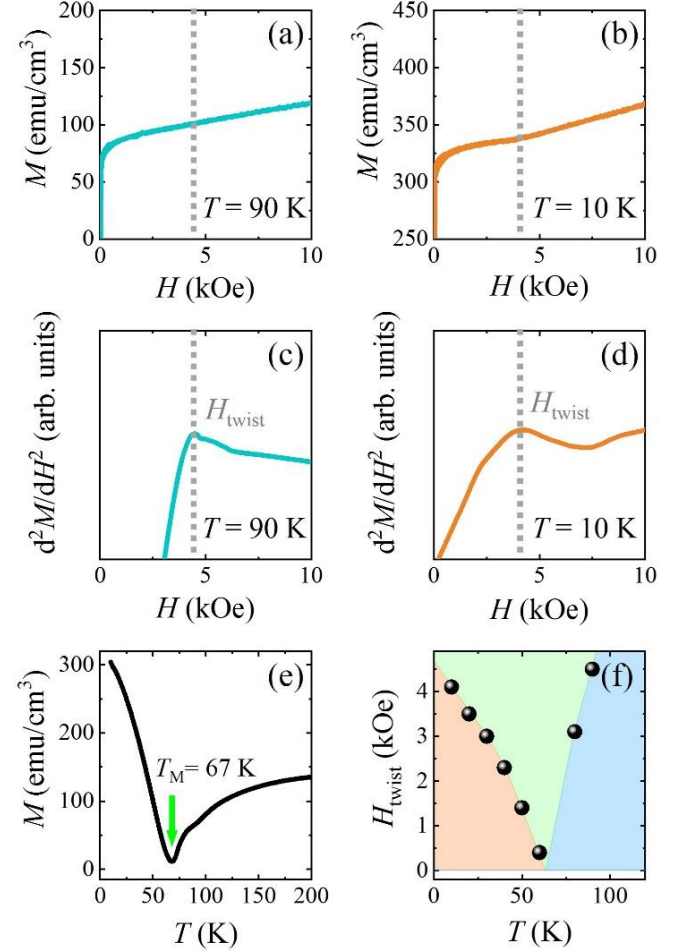

Fig. 2. The positive half branches of the magnetization curves at (a) $T = 90$ K and (b) $T = 10$ K. The second derivative of the magnetization curves ($d^2M/dH^2$) at (c) $T = 90$ K and (d) $T = 10$ K present the accurate twisting field $H_{twist}$ (4.5 kOe for $T = 90$ K and 4.1 kOe for $T = 10$ K). The grey dot lines mark the $H_{twist}$ in the magnetization and $d^2M/dH^2$ curves. (e) The temperature-dependent in-plane magnetization at $H = 50$ Oe; the green arrow indicates the compensation temperature $T_M$. (f) The phase diagram of the Py/Gd multilayer is illustrated in terms of $H_{twist}$, i.e. Gd-aligned phase (red region), twisted state (green region), and Py-aligned phase (blue region).

The conventional method for identifying the twisted state is to measure the bending of the magnetization curves [15]. To address the accurate $H_{twist}$ of the twisted state, the second derivative of the magnetization curves

**3 / 11**

($d^2M/dH^2$) is presented in Fig. 2 (c) and (d) [21]. The bumps observed in the $d^2M/dH^2$ curves are the precise measures of $H_{twist}$, marked by the grey dotted lines in Fig. 2 (c) and (d). $H_{twist}$ is 4.5 kOe for $T$ = 90 K and 4.1 kOe for $T$ = 10 K, respectively. Then the phase diagram of the Py/Gd multilayer can be depicted according to the temperature dependence of $H_{twist}$ [Fig. 2(f)]. $M_{Py}$ governs the macroscopic magnetization for $T$ > 67 K and $M_{Gd}$ becomes dominant for $T$ < 67 K. A sufficiently strong magnetic field ($H > H_{twist}$) can introduce the twisted state in the Py/Gd multilayer. $H_{twist}$ is weak in the vicinity of $T_M$ (67 K) and rises rapidly when changing the temperature away from $T_M$. Note that the increase in $H_{twist}$ is steeper for $T$ > 67 K than for T < 67 K. Overall, figure 2 presents a magnetic characterization of the Py/Gd multilayer following the conventional method, and figure 2 (f) depicts a well-known phase diagram for such artificial ferrimagnets [20].

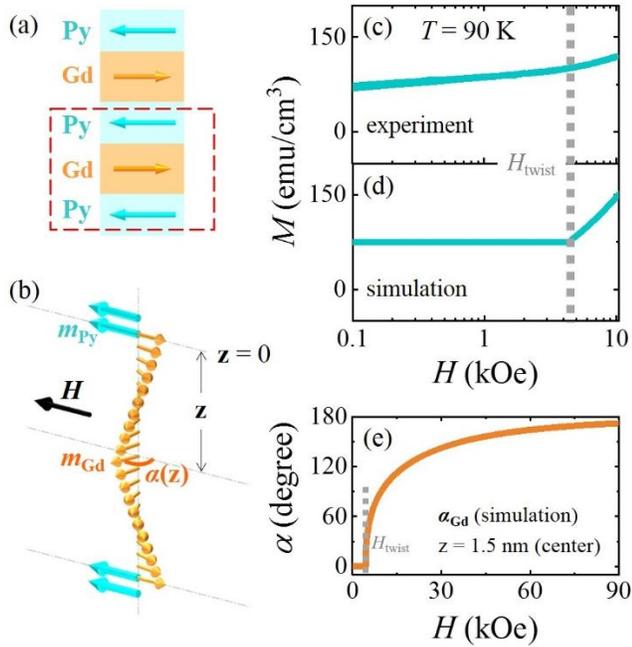

Fig. 3. Schematics of (a) the Py/Gd multilayer and (b) the magnetic winding within the Gd layer (thickness is 3 nm) for $M_{Py} > M_{Gd}$. The magnetic winding can be quantitatively characterized by a winding angle $\alpha_{Gd}$. The magnetization curves at $T$ = 90 K obtained (c) in the experiment and (d) in the simulation. The simulation well reproduces the $H_{twist}$ in the magnetization curve. (e) The $H$-dependent $\alpha_{Gd}$ of the center Gd moment (depth $z$ = 1.5 nm) quantitatively depicts the magnetic winding inside the Gd layer. The grey dot lines mark the $H_{twist}$ in the magnetization and $\alpha_{Gd}$ curves in (c), (d) and (e).

Although the phase diagram of artificial ferrimagnets is accessible via the conventional magnetization-curve measurement, the depth-dependent non-collinear magnetic profile in the twisted state is inaccessible using this method. To reveal the magnetic profile of the twisted state, we study the magnetization evolution of the Py/Gd multilayer in an in-plane $H$ by the micromagnetic simulation. The Py/Gd multilayer is modeled into three regions, i.e. the Py layer, the Gd layer, and the Py/Gd interface, respectively. The magnetic parameters of these three regions can be quoted from the literature and our previous work [13,19]. Exchange stiffness are $A_{Py} = 1\times10^{-6}$ erg/cm in the Py layer and $A_{Gd} = 0.35\times10^{-7}$ erg/cm in the Gd layer, the interfacial exchange stiffness is $A_{int} = -8\times10^{-7}$ erg/cm for the antiferromagnetic coupling at the Py/Gd interface.

We first simulate the magnetization curve at $T$ = 90 K with the magnetization $M_{Py}$ = 660 emu/cm$^3$ and $M_{Gd}$ = 590 emu/cm$^3$. The magnetization are retrieved according to the macroscopic magnetization of the multilayer [Fig. 2 (e)] and a Py reference sample, which will be discussed later in Fig. 7 [22]. The micromagnetic simulation shows a magnetic twisting in the Py/Gd multilayer when $H$ exceeds a threshold field. Since the Gd is a weaker ferromagnet than the Py ($A_{Gd} \ll A_{Py}$), the magnetic twisting mainly occurs within the Gd layer when the Py layer remains as a single domain. The in-plane $H$ drags the inner Gd moment deviating from the original direction, meanwhile, the interfacial Gd moment at the Py/Gd interface is pinned by the Py layer. As the result, the linear rotation of the local Gd moment introduce a magnetic winding within the Gd layer when $M_{Py}$ aligns with $H$ firmly [Fig. 3 (b)]. Such magnetic winding inside the Gd layer can be characterized by a winding angle $\alpha_{Gd}$. In our Py/Gd multilayer, two Gd layers are sandwiched by three Py layers so that the multilayer possesses a mirror symmetry along the film normal direction. Hence the winding angle $\alpha_{Gd}$ inside two Gd layers are identical. The simulation reproduce the magnetization curve at $T$ = 90 K with $H_{twist}$ in good agreement with the experimental data [Fig. 3 (c) and (d)]. The depth-dependent magnetic profile in the twisted state is also accessible in the simulation. Figure 3 (e) presents the $H$-dependent $\alpha_{Gd}$ of the center Gd moment (depth $z$ = 1.5 nm in the Gd layer), whereby the magnetic winding inside the Gd layer is quantitatively depicted. Here the simulations oversimplify the interfacial antiferromagnetic coupling in the Py/Gd multilayer, which can be improved by a more complicated model [19]. Nevertheless, the micromagnetic simulations well reproduce the $H_{\text{twist}}$ in the magnetization curves and reveal the non-collinear magnetic profile in the twisted state, which provides a quantitative portrait of the twisted state that cannot be directly acquired from the conventional magnetization-curve measurement.

In contrast to the conventional magnetization-curve measurement, AMR is usually more sensitive to non-collinear magnetic structures [23]. Figure 4 presents the AMR measurements of the Py/Gd multilayer at $T$ = 90 K, aiming to quantitatively characterize the twisted state. The electrical current ($I$ = 0.1 mA) flows along the main stripe of the Hall bar. $H$ can be swept along any azimuthal angle $\theta_H$ in the film plane [Fig. 4(a)]. Figure 4 (b) shows the $H$-dependent longitudinal resistance $R(H)$ for $H$ // $I$ ($\theta_H = 0°$) and $H \perp I$ ($\theta_H = 90°$). Around zero $H$, $R_{H\|I}(H)$ is higher than $R_{H\perp I}(H)$ and a long tail can be observed in both geometries, which is not the typical AMR signal during the magnetization reversal. If the magnetic twisting occurs within a narrow thickness range (a few nanometers) in the multilayer, there could be two different contributions to the magnetoresistance. In addition to the AMR, such non-collinear magnetic structure can cause the spin-dependent scattering, which is a GMR-type magnetoresistance (independent of the orientation of the magnetic moments with respect to the current direction) [24]. To distinguish between the AMR and GMR-type magnetoresistance, we calculate the differential value of $R(H)$ as $R_{diff}(H) = R_{H\|I}(H) - R_{H\perp I}(H)$ [Fig. 4 (d)]. Given the harmonic $(\cos\theta_H)^2$ dependence of the AMR signal, this procedure can readily eliminate GMR-type magnetoresistance, leaving the AMR signal that depends on the angle between the magnetic moment and current [25]. As shown in Fig. 4 (d), $R_{diff}(H)$ remains a constant below a threshold field. Above this threshold field, a sharp drop in $R_{diff}(H)$ is observed, accompanied with a gradual upturn. This gradual upturn is unsaturated even at $H$ = 90 kOe, leading to a ⌞-shaped $R_{diff}(H)$ curve. Here the threshold field is about 4.5 kOe, in consistent with $H_{\text{twist}}$ at $T$ = 90 K. Hence the ⌞-shaped $R_{diff}(H)$ curve is associated with the twisted state of the Py/Gd multilayer, i.e. originating from the magnetic winding within the Gd layer.

The AMR signals are also recorded in a rotating $H$ of various strengths [Fig. 4 (c)]. The differential magnitude of AMR can be calculated using the AMR signals of $\theta_H =$ 0° and $\theta_H = 90°$ [$AMR(0°)$ - $AMR(90°)$]. As shown in Fig. 4 (d), the excellent consistency between $R_{diff}(H)$ and $AMR(0°)$ - $AMR(90°)$ confirm the source of the $R_{diff}(H)$ being the AMR effect.

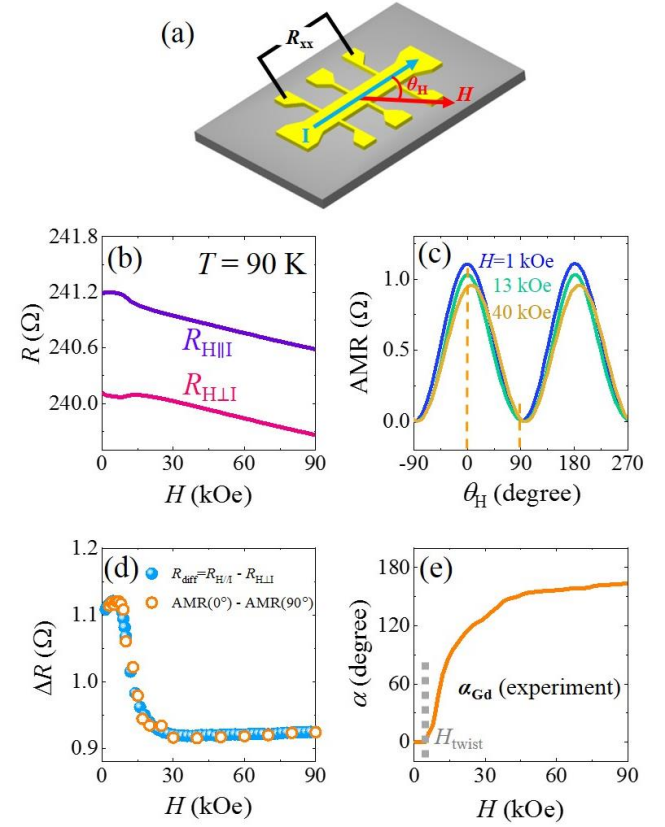

Fig. 4. (a) Schematic of the AMR measurements. (b) The longitudinal resistance for $H$ // $I$ ($\theta_H = 0°$) and $H \perp I$ ($\theta_H = 90°$) at $T$ = 90 K. (c) The AMR signals in a rotating $H$ of 1 kOe, 13 kOe and 40 kOe ($T$ = 90 K). (d) The $H$-dependent differential resistance $R_{diff}(H)$ and differential AMR results obtained from [ $AMR(0°)$ - $AMR(90°)$]. (e) The winding angle $\alpha_{\text{Gd}}$ within the Gd layer is obtained according to the $H$-dependent $R_{diff}(H)$ and AMR results in (d), utilizing Eqns. (2) and (3). The grey dot line indicates the $H_{\text{twist}}$ in the $\alpha_{\text{Gd}}$ curve in (e).

In the twisted state of the Py/Gd multilayer, the AMR signals can be estimated as the parallel circuits of the local resistivity [26,27], with the magnetization orientation $\alpha(z)$ distributed as a function of the depth $z$ (denoted in Fig. 3 (b)). Note that AMR is an even function of the magnetization orientation $\alpha(z)$ with respect to the current direction, i.e. $\rho(\alpha) = \rho_\perp + \Delta\rho \cos^2\alpha$ [28]. Here $\rho_\perp$ is the resistivity for $\alpha = 90°$ and $\Delta\rho$ is the magnitude of the resistivity change due to AMR. Therefore, the local resistivity due to AMR is a depth-dependent resistivity $\rho(z)$ and the sheet resistance $R$ of the multilayer is given by [29]

$$\frac{1}{R} = \int \frac{dz}{\rho(z)} = \int \frac{dz}{\rho_\perp + \Delta\rho \cos^2[\alpha(z)]}. \quad (1)$$

Here $z$ denotes the depth inside the Gd layer from the interface [Fig. 3 (b)]. In the transformation from the Py-aligned phase into the twisted state, $M_{Py}$ aligns with $H$ firmly so that the variation in the AMR of the Py layer is negligible. Both the twisted state and $\rho(z)$ depend on the winding angle $\alpha_{Gd}$ within the Gd layer. Assuming a uniform twist of the Gd moments [12], the sheet resistances of the Gd layer ($R_{Gd}$) and the Py/Gd multilayer ($R_{Py/Gd}$) can be calculated using Eqn. (1) by follows:

$$\frac{1}{R_{Gd}} = \frac{d_{Gd} \cdot \tan^{-1}\left(\sqrt{\frac{\rho_\perp^{Gd}}{\rho_\parallel^{Gd}}} \cdot \tan\alpha_{Gd}\right)}{\sqrt{\rho_\perp^{Gd}\rho_\parallel^{Gd}} \cdot \alpha_{Gd}}, \quad (2)$$

$$\frac{1}{R_{Py/Gd}} = \frac{2}{R_{Gd}} + \frac{3}{R_{Py}}, \quad (3)$$

where $d_{Gd}$ is the thickness of the Gd layer. $\rho_\perp^{Gd}$ and $\rho_\parallel^{Gd}$ are the resistivity of the Gd layer for $\alpha_{Gd} = 90°$ and $\alpha_{Gd} = 0°$ (see supplementary Fig. S2 (b)) [30]. $R_{Py}$ is the resistance of the Py layer (see supplementary Fig. S2 and Fig. S3) [30]. Equations (2) and (3) allow us to calculate $R_{Py/Gd}$ if the winding angle $\alpha_{Gd}$ is known, and vice versa. Based on the $R_{diff}(H)$ curve presented in Fig. 4 (d), we can acquire the $\alpha_{Gd}$ as a function of $H$ directly using Eqns. (2) and (3). The $H$-dependent $\alpha_{Gd}$ shows a good agreement with the simulation result [Fig. 3 (e) versus Fig. 4 (e)], further validating the fact that the twisted state is caused by the magnetic winding within the Gd layer when $M_{Py} > M_{Gd}$. In particular, the $H$-dependent $\alpha_{Gd}$ produces a ∟-shaped $R_{diff}(H)$ curve whereby the $\alpha_{Gd}$ can be quantitatively determined in the AMR experiment.

Owing to the mirror symmetry of the Py/Gd multilayer, we originally expected the similar AMR results for the scenarios of $M_{Py} > M_{Gd}$ and $M_{Py} < M_{Gd}$. However, as shown in Fig. 5 (a), $R_{H\parallel I}(H)$ and $R_{H\perp I}(H)$ recorded at $T = 10$ K exhibit the distinct features with respect to their counterparts at $T = 90$ K. A crossover between $R_{H\parallel I}(H)$ and $R_{H\perp I}(H)$ curves is observed at $T = 10$ K. Figure 5 (c) plots the $R_{diff}(H)$ curve at $T = 10$ K, confirming the distinct features of the AMR signals at $T = 10$ K. A sharp drop above a threshold field ($H_{twist1}$) is observed in the $R_{diff}(H)$ curve at $T = 10$ K, followed by an upturn. Then the second drop ensues at a much higher threshold field ($H_{twist2}$) than $H_{twist1}$, accompanied by a sign change at $H \sim 45$ kOe. Subsequently, the $R_{diff}(H)$ curve reaches a negative maximum at $H \sim 65$ kOe, followed by the second upturn. This sign change at $H \sim 45$ kOe is also visible in the AMR signals in Fig. 5 (b), confirming that two sequential drops in the $R_{diff}(H)$ curve are unambiguously attributed to the AMR effect. The sign change at $H \sim 45$ kOe is due to the horizontal shift of the harmonic AMR signals [19]. As validated in Fig. 4, the twisted state of the Py/Gd multilayer (with the twisting field $H_{twist}$ as a feature) results in a sharp drop in the $R_{diff}(H)$ curve. Therefore, two sequential drops in the $R_{diff}(H)$ curve signify two distinct twisted states with different twisting fields ($H_{twist1}$ and $H_{twist2}$) at $T = 10$ K [Fig. 5 (c)]. However, only one bump at $H_{twist1}$ is observed in the $d^2M/dH^2$ curve at $T = 10$ K [Fig. 5 (d)], which is the signature of the first twisted state. The second twisted state is invisible in the $d^2M/dH^2$ curve.

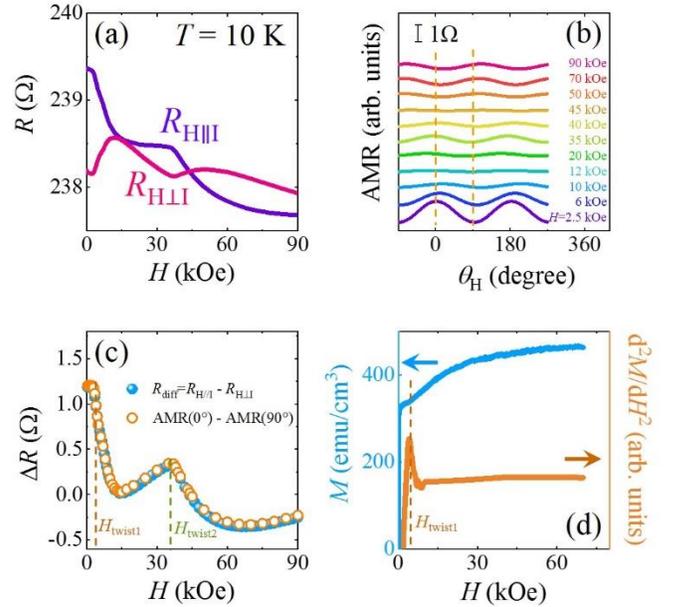

Fig. 5. (a) The longitudinal resistance for $H // I$ and $H \perp I$ at $T = 10$ K. (b) The AMR signals in a rotating $H$ of various strengths at $T = 10$ K. (c) The $H$-dependent differential resistance $R_{diff}(H)$ and differential AMR results obtained from $[AMR(0°) - AMR(90°)]$. (d) $M$-$H$ and $d^2M/dH^2$ curves at $T = 10$ K. The second twisting field $H_{twist2}$ observed in $R_{diff}(H)$ curve is invisible in the $d^2M/dH^2$ curve.

To reveal the non-collinear magnetic profiles of two twisted states, we perform further micromagnetic simulations for the scenario of $M_{Py} < M_{Gd}$. When the in-plane $H$ is weak in such scenario, $M_{Py}$ is originally antiparallel to $H$ and $M_{Gd}$ is aligned with $H$ firmly. If $H$ is sufficiently strong, the Zeeman energy in the Py layer would become competitive with the exchange energy in the Gd layer. Given the rigid antiferromagnetic coupling at the Py/Gd interface ($A_{int} \gg A_{Gd}$) and $A_{Py} \gg A_{Gd}$ [16, 31], $M_{Py}$ could deviate from the original direction and introduce a magnetic winding inside the Gd layer. Note that the inner Py layer in the multilayer (Py-center in Fig. 6 (a)) possesses the mirror symmetry while the outermost Py layer (Py-top and Py-bottom in Fig. 6 (a)) does not, the threshold fields for $M_{Py}$ reorientation should be different in the Py-top and Py-center layers. The $M_{Py}$ rotation in the Py-top layer causes the magnetic winding within one of the Gd layers in the Py/Gd multilayer [Fig. 6 (b)]. In contrast, the $M_{Py}$ rotation in the Py-center layer causes magnetic winding in both Gd layers [Fig. 6 (c)], resulting in a much steeper increase in Gd exchange energy, i.e., a much higher threshold field. Namely, for the scenario of $M_{Py} < M_{Gd}$, the broken mirror symmetry of the outermost Py layer causes the emergence of the second twisted state with $H_{twist1}$.

Figure 6 (d) plots the simulation results of the $M_{Py}$ rotation angles in the Py-top and Py-center layers ($\alpha_{Py\text{-}top}$ and $\alpha_{Py\text{-}center}$ in Fig. 6 (b) and (c)). The simulation parameters are the following: exchange stiffness $A_{Py} = 1\times10^{-6}$ erg/cm, $A_{Gd} = 0.9\times10^{-7}$ erg/cm and $A_{int} = -8\times10^{-7}$ erg/cm; the magnetization $M_{Py} = 660$ emu/cm$^3$ and $M_{Gd} = 1500$ emu/cm$^3$. The mean field theory suggests that exchange stiffness scales linearly with the magnetization ($A(T) \propto M(T)$) [31,32], so we retain the linear dependence of $A_{Gd}$ on $M_{Gd}$ in the simulations. A larger $M_{Gd}$ indicates a larger $A_{Gd}$ at $T = 10$ K compared to the counterparts at $T = 90$ K. The simulation results show the completely different trajectories of $\alpha_{Py\text{-}top}$ and $\alpha_{Py\text{-}center}$ against $H$. A sharp increase in $\alpha_{Py\text{-}top}$ is observed at $H_{twist1}$, while the increase in $\alpha_{Py\text{-}center}$ occurs at $H_{twist2}$. Owing to the tiny $\alpha_{Py\text{-}top}$ in the vicinity of $H_{twist1}$, the Gd exchange energy is negligible with respect to the Zeeman energy in the Py layers around $H_{twist1}$. Hence a small variation of $\alpha_{Py\text{-}center}$ around $H_{twist1}$ may also develop in conjunction with the increase of $\alpha_{Py\text{-}top}$. Such a small variation of $\alpha_{Py\text{-}center}$ is instantaneously suppressed by the onset of magnetic winding inside the Gd layer and therefore has little effect on the macroscopic magnetization and AMR result. Both $\alpha_{Py\text{-}top}$ and $\alpha_{Py\text{-}center}$ curves are unsaturated at $H = 90$ kOe, which in principle should give rise to an unsaturated AMR signal at $T = 10$ K [Fig. 5 (c)].

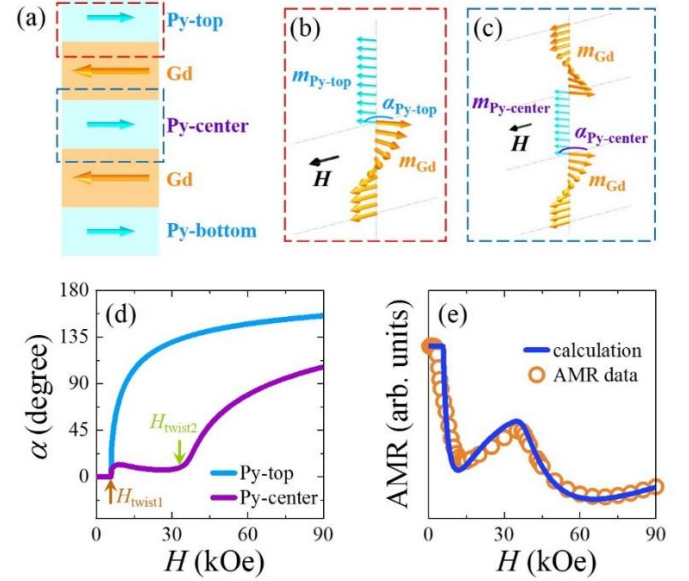

Fig. 6. (a) Schematic of the Py/Gd multilayer for $M_{Py} < M_{Gd}$. Schemes of the $M_{Py}$ rotation (b) in the Py-top layer and (c) in the Py-center layer, associated with the magnetic winding within the Gd layers for $M_{Py} < M_{Gd}$. The $M_{Py}$ rotation can be quantitatively characterized by $\alpha_{Py\text{-}top}$ and $\alpha_{Py\text{-}center}$. (d) The $M_{Py}$ rotation angles in the Py-top and Py-center layers ($\alpha_{Py\text{-}top}$ and $\alpha_{Py\text{-}center}$) obtained in the simulations. The onsets of $\alpha_{Py\text{-}top}$ and $\alpha_{Py\text{-}center}$ occur at $H_{twist1}$ and $H_{twist2}$ respectively. (e) The calculated AMR curve using $\alpha_{Py\text{-}top}$ and $\alpha_{Py\text{-}center}$ reproduces the trajectory of the AMR data at $T = 10$ K.

Utilizing the $H$-dependent $\alpha_{Py\text{-}top}$ and $\alpha_{Py\text{-}center}$ curves, as well as the depth-dependent $\alpha_{Gd}$ within the Gd layer (see supplementary Fig. S4) [30], we can calculate the AMR curve using Eqn. (1) directly instead of Eqn. (2). $\rho_\perp$ and $\Delta\rho$ can be retrieved in supplementary S2 [30]. The calculation result shown in Fig. 6 (e) perfectly reproduces the trajectory of the AMR data at $T = 10$ K, further confirming the fact that two twisted states with $H_{twist1}$ and $H_{twist2}$ emerge from the $M_{Py}$ rotation in the outermost and inner Py layers, respectively. In other words, the $M_{Py}$ rotation in the outermost and inner Py layers can be monitored in terms of $H_{twist1}$ and $H_{twist2}$.

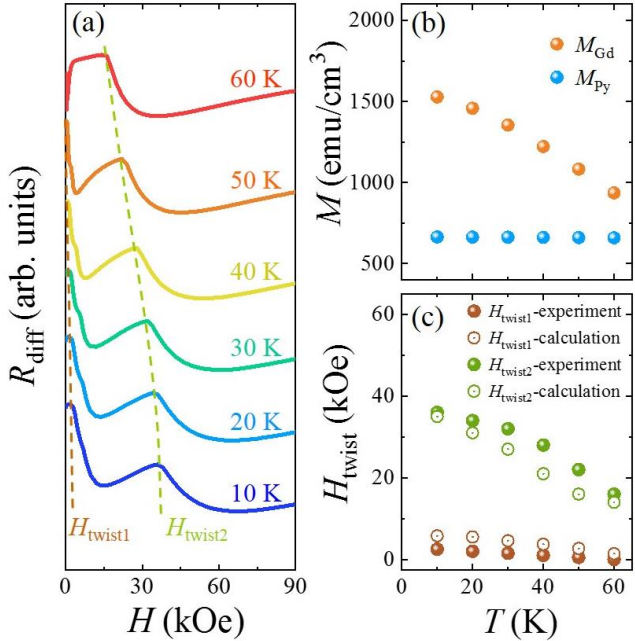

Fig. 7. (a) The AMR curves in the temperature range from 10 K to 60 K. Two dashed lines mark $H_{twist1}$ and $H_{twist2}$ in the AMR curves. (b) The temperature-dependent $M_{Py}$ and $M_{Gd}$. (c) $H_{twist1}$ and $H_{twist2}$ obtained in the AMR measurements and calculations. The good agreement between the experimental and calculation results confirms that the temperature-dependent $H_{twist1}$ and $H_{twist2}$ are due to the temperature-dependent $M_{Gd}$.

Figure 7 (a) plots the AMR curves in the temperature range from 10 K to 60 K. $H_{twist1}$ and $H_{twist2}$ are extracted and shown in Fig. 7 (c). Both $H_{twist1}$ and $H_{twist2}$ decrease monotonically as the temperature approaches $T_M$. Given the temperature-dependent macroscopic magnetization of the Py/Gd multilayer [Fig. 2 (e)] and the temperature-dependent $M_{Py}$ of a Py reference sample [Fig. 7 (b)], the temperature-dependent $M_{Gd}$ can be readily deduced [33]. Taking the temperature-dependent $M_{Gd}$ into account, we can calculate $H_{twist1}$ and $H_{twist2}$ following the same procedures as in Fig. 6 (d) and (e). The linear dependence of $A_{Gd}$ on $M_{Gd}$ is retained in the calculations. The calculated $H_{twist1}$ and $H_{twist2}$ well reproduce the monotonic trajectories of the experimental data. The good agreement between the calculation and experimental results provides an explicit evidence that the temperature-dependent $H_{twist1}$ and $H_{twist2}$ are explainable in the scope of the temperature-dependent $M_{Gd}$.

As mentioned above, the Gd is a weak ferromagnet ($A_{Gd} \ll A_{Py}$ and $A_{Gd} \ll A_{int}$) so that the magnetic winding occurs mainly within the Gd layer. Given the linear dependence of $A_{Gd}$ on $M_{Gd}$, the reduced $M_{Gd}$ corresponds to a reduced $A_{Gd}$ and a reduction of the Gd exchange energy due to the magnetic winding. As illustrated in Fig. 6 (b) and (c), two twisted states are both correlated with the magnetic winding inside the Gd layers. In consequence, $A_{Gd}$ determines the energy of the magnetic winding and thus the $H_{twist1}$ and $H_{twist2}$ of two twisted states. Namely, the reduced $M_{Gd}$ leads to a decrease in $H_{twist1}$ and $H_{twist2}$ during the temperature rise, as shown in Fig. 7 (a).

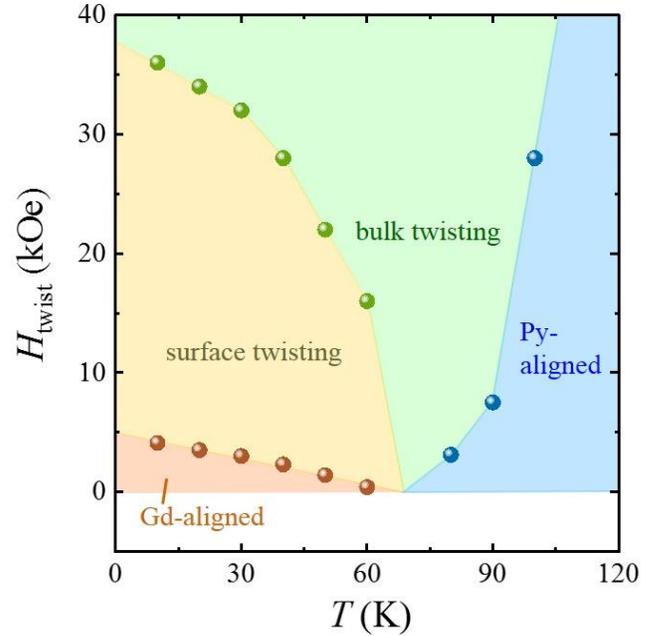

Fig. 8. The updated phase diagram describes the comprehensive magnetic structure in the Py(2.5 nm)/Gd(3 nm)/Py(2.5 nm)/Gd(3 nm)/Py(2.5 nm) multilayer. Instead of a unitary twisted state, the updated phase diagram highlights two regimes of surface twisting and bulk twisting.

According to the temperature-dependent $H_{twist1}$ and $H_{twist2}$ for $T < T_M$ and the temperature-dependent $H_{twist}$ for $T > T_M$, the phase diagram of the Py/Gd multilayer is presented in Fig. 8. The updated phase diagram differs from that shown in Fig. 2 (f) wherein the twisted state is divided into two regimes. In the scenario of $M_{Gd} > M_{Py}$, the regime between $H_{twist1}$ and $H_{twist2}$ corresponds to the $M_{Py}$ rotation in the outermost Py layers accompanied by the magnetic winding inside the neighboring Gd layers, which could be referred to as surface twisting. The second regime above $H_{twist2}$ addresses the $M_{Py}$ rotation in the inner Py layer and the magnetic winding within the inner Gd layers, referring to as bulk twisting. In the scenario of $M_{Gd} < M_{Py}$, the magnetic winding within the inner Gd layers can be launched above $H_{twist}$, which can also be called bulk twisting. Namely, the scenario of $M_{Gd} < M_{Py}$ permits only bulk twisting while both surface twisting and bulk twisting are allowed for $M_{Gd} > M_{Py}$. In both scenarios, both Py and Gd moments are involved in the magnetic twisting. Therefore, the terms "surface twisting" and "bulk twisting" are used instead of "surface Py twisting" and "bulk Py twisting" [15]. The updated phase diagram depicts a comprehensive portrait of the complex magnetic structure in the Py/Gd multilayer, which is also valid in other artificial ferrimagnets. The construction of this phase diagram may facilitate the understanding of the chiral magnon modes [13,34] and the depth-dependent magnetic profile of magnetic skyrmions [35] in artificial ferrimagnets.

## IV. DISCUSSION

It is worth noting that the conventional magnetization-curve measurement picks up the macroscopic magnetization projection along the external $H$, which is proportional to $M_{net} \cos\alpha$. The Py/Gd multilayer is a compensation magnetization system where $M_{Py}$ and $M_{Gd}$ compensate each other, leading to a tiny $M_{net}$ in comparison to $M_{Py}$ and $M_{Gd}$. Meanwhile, $\cos\alpha$ is a less sensitive function of $\alpha$ in comparison to $\cos^2\alpha$. Thus the bulk twisting makes a negligible contribution to the $d^2M/dH^2$ curve and remains invisible in the conventional magnetization-curve measurement. In contrast, the outermost $M_{Py}$ in the multilayer is uncompensated, leading to a detectable signal in $d^2M/dH^2$ curve (i.e. surface twisting).

In contrast, AMR signal is proportional to $\Delta\rho \cdot \cos^2\alpha$ (i.e. two-fold symmetry). Thus the AMR contributions of Py and Gd are constructive rather than destructive to each other in spite of the antiferromagnetic coupling between $M_{Py}$ and $M_{Gd}$ in the multilayer [36]. As the result, an enlarged $\Delta\rho$ can be observed in the Py/Gd multilayer compared to a single Py or Gd layer. In addition, $\cos^2\alpha$ is a sensitive function of $\alpha$ with respect to $\cos\alpha$. Therefore, AMR is an ideal probe of the non-collinear magnetic structure in artificial ferrimagnets. Both surface twisting and bulk twisting can be characterized by the AMR measurement.

The updated phase diagram is not symmetric for $M_{Gd} < M_{Py}$ and $M_{Gd} > M_{Py}$. Only bulk twisting is allowed when the outermost layer in the multilayer dominates the macroscopic magnetization, e.g., the outermost Py layer is the master in the scenario of $M_{Gd} < M_{Py}$. In contrast, both surface twisting and bulk twisting are allowed if the outermost layer is the slave, e.g., the outermost Py layer in the scenario of $M_{Gd} > M_{Py}$. In another word, the emergence of surface twisting is due to the broken mirror symmetry of the slave layer. To further confirm this statement, the AMR curves of the Gd/Py/Gd/Py/Gd multilayer are calculated in the scenario of $M_{Gd} < M_{Py}$ (see supplementary Fig. S5) [30]. In this case, two Py layers are the master and are aligned to the external $H$. The outermost Gd layer has only one Gd/Py interface and starts to form a magnetic twisting at a small threshold field. On the contrary, the center Gd layer is sandwiched by two Py layers so that the magnetic twisting occurs at a much higher threshold field. Therefore, two distinct twisted states can be observed in the Gd/Py/Gd/Py/Gd multilayer at $T > T_M$, as the counterparts observed in the Py/Gd/Py/Gd/Py multilayer at $T < T_M$.

The AMR works excellently as a probe for the twisted states in our Py/Gd multilayer sample. Then it is an appealing question how AMR works for Py/Gd multilayers with larger repetition numbers. Using the micromagnetic simulation and the same procedures as in Fig. 6 (d) and (e), we also calculate the AMR curves with the magnetic parameters at $T = 10$ K (see supplementary S1) [30]. The repetition numbers are 5 layers, 9 layers, 21 layers, and 51 layers in the respective simulations. The calculation results clearly show the repetition-number dependent AMR curves (see supplementary Fig. S1) [30]. As the repetition number increases, the surface twisting in the multilayer contributes to a smaller drop in the AMR signal, in conjunction with a greater contribution of bulk twisting. Therefore, the AMR signal of surface twisting would become invisible when the repetition number is sufficiently high, and bulk twisting would dominate the change of AMR signal. In other words, we cannot detect the surface twisting in the Py/Gd multilayer by AMR measurements if the repetition number is too high.

## V. SUMMARY

In summary, the twisted state of the artificial ferrimagnetic Py/Gd multilayer was systematically investigated, with surface twisting and bulk twisting quantitatively revealed by AMR measurements. On the contrary, the conventional magnetization-curve measurement only revealed the surface twisting. Our results manifest AMR as an ideal probe of the non-collinear magnetic structure in artificial ferrimagnets. The updated phase diagram may facilitate the study of the chiral magnon modes and magnetic skyrmions in artificial ferrimagnets.

## ACKNOWLEDGMENTS

This work is supported by the National Key Research and Development Program of China (Nos. 2022YFA1405100), the National Natural Science Foundation of China (No. 11874072, No.11874416, No. 11974408), and the Strategic Priority Research Program of the Chinese Academy of Sciences (No. XDB33020300).